\documentclass[11pt,fleqn]{article}

\usepackage{amsmath,amssymb,amsthm,enumerate}%,backref,cite

\allowdisplaybreaks

\newcommand{\ds}{\displaystyle}
\newcommand{\const}{\mathop{\rm const}\nolimits}

\newcommand{\todo}[1][\null]{\ensuremath{\clubsuit}}

\newcommand{\noprint}[1]{}

\newcounter{tbn}

\newcounter{mcasenum}

{\theoremstyle{definition}

}

\begin{document}

\par\noindent {\LARGE\bf
Reduction operators and exact solutions \\ of generalized Burgers equations
\par}

{\vspace{4mm}\par\noindent {\large O.A. Pocheketa$^\dag$, R.O. Popovych$^\ddag$}
\par\vspace{2mm}\par}

{\vspace{2mm}\par\noindent \it
$^\dag{}^\ddag$Institute of Mathematics of NAS of Ukraine, 3 Tereshchenkivska Str.,
01601 Kyiv, Ukraine\par}

{\vspace{2mm}\par\noindent {\it $^\ddag\phantom{^\dag}$Wolfgang Pauli Institute, Nordbergstra{\ss}e 15, A-1090 Wien,
Austria}\par} {\vspace{2mm}\noindent {\it $\phantom{^\dag{}^\ddag}${\rm E-mail:}
$^\dag$pocheketa@yandex.ru, $^\ddag$rop@imath.kiev.ua }\par}

{\vspace{7mm}\par\noindent\hspace*{8mm}\parbox{144mm}{\small
Reduction operators of generalized Burgers equations are studied.
A connection between these equations and potential fast diffusion equations with power nonlinearity $-1$ via reduction operators is established.
Exact solutions of generalized Burgers equations are constructed using this connection 
and known solutions of the constant-coefficient potential fast diffusion equation.
}\par\vspace{2mm}}

\section{Introduction}

In order to construct exact solutions of a partial differential
equation, techniques involving nonclassical symmetries (also called conditional
symmetries, $Q$-conditional symmetries or reduction operators) are used, which
leads to ansatzes reducing the initial equation to another one with a less number of
independent variables. The nonclassical reduction method was proposed by Bluman and
Cole \cite{blum68th,blum69a} and applied to many models of real-world phenomena,
see~\cite{fush93book,olve96book} and references therein. We use the name
``reduction operator'' for the notion of nonclassical symmetries, which is justified
by results of~\cite{zhda99a}.

The classical and generalized Burgers equations are used to model a wide variety of
phenomena in physics, chemistry, mathematical biology, etc., see, e.g., \cite[Chapter~4]{whit74book}.
In the present paper we study the class of generalized Burgers equations of the form
\begin{equation}\label{GBE}
u_t+uu_x+f(t,x)u_{xx}=0
\end{equation}
where the arbitrary element~$f$ runs through the set of nonvanishing smooth functions of~$t$ and~$x$. 
These equations were intensively investigated within the framework of symmetry analysis. 
The maximal Lie invariance group of the
classical Burgers equation (with $f=1$) was computed in~\cite{katk65a}. It is the
Burgers equation that was first considered from the nonclassical symmetry point of
view after the seminal paper~\cite{blum69a}. The corresponding results obtained
in~\cite{wood71th,wood71a} can be found in~\cite{ames72book}.
After a longtime break the interest to symmetry analysis of various generalizations of the Burgers equation was renewed.
A point transformation between equations of the form~\eqref{GBE} with~$f=f(t)$ was detected in~\cite{cate89a}.
The study of form-preserving (admissible) transformations for class~\eqref{GBE} by Kingston and Sophocleous~\cite{king91c}
was a pioneer work on form-preserving transformations in the literature.

After~\cite{ames72book,wood71th,wood71a}, nonclassical symmetries of the Burgers equation were considered in a series of papers.
In~\cite{pucc92a} solving the corresponding determining equations was partitioned into three cases,
$\xi_u=0$, $\xi_u=1$ and $\xi_u=-\frac12$, assuming $\tau=1$ (cf.\ Section~\ref{SectionOnDetEqs}).
It was proved that each nonclassical symmetry with $\xi_u=0$ is equivalent to a Lie symmetry.
The case $\xi_u=1$ is rather simple and gives a single nonclassical symmetry.
Particular solutions of the determining equations were found in the last case.
Analogous results were presented in~\cite{arri93a}.
Nonclassical symmetries of a system of differential equations that is equivalent to the Burgers equation
were studied in~\cite{olve96book} in a similar way.
The ``no-go'' case $\xi_u=-\frac12$ was studied in~\cite{mans99a} and then in~\cite{arri02a}.

\looseness=-1
Group classification of the subclass of the
class~(1) singled out by the constraint $f_x=0$ was carried out in~\cite{wafo04d}.
It was also shown that nontrivial nonclassical symmetries, which are not equivalent to Lie symmetries, exist only for $f=\const$,
i.e.\ for the classical Burgers equation.
Reduction operators of equations from the entire class~\eqref{GBE} have not been exhaustively studied yet.

In the present paper we investigate reduction operators of generalized Burgers equations of the form~\eqref{GBE}
and establish their connection with solutions of potential fast diffusion equations with power nonlinearity $-1$.
Known exact solutions of the constant-coefficient potential fast diffusion equation allow us to
construct parameterized families of exact solutions
of certain generalized Burgers equations from the class~\eqref{GBE}.

The structure of this paper is as follows.
The equivalence group of the class~\eqref{GBE} is presented in the next section.
Ibid the determining equations for the coefficients of reduction operators are derived and preliminary studied.
Section~\ref{MAINsection} reveals the above connection with potential fast diffusion equations.
In Section~\ref{LASTsection} we construct exact solutions of equations from the class~\eqref{GBE}.

\section{Determining equations}\label{SectionOnDetEqs}

The class~\eqref{GBE} is normalized, i.e.\ any admissible transformation of this
class is generated by a transformation from its equivalence group
(see~\cite{fush92,popo10a} for precise definitions). The equivalence group $G^\sim$
of the class~\eqref{GBE} consists of the transformations
\begin{equation*}
\tilde{t}=\frac{\alpha t+\beta}{\gamma t+\delta},\;\;\tilde{x}=\frac{\kappa x+\mu_1
t +\mu_0}{\gamma t+\delta},\;\;\tilde{u}=\frac{\kappa(\gamma t+\delta)u-\kappa\gamma
x+\mu_1\delta-\mu_0\gamma}{\alpha\delta-\beta\gamma},\;\;
\tilde{f}=\frac{\kappa^2}{\alpha\delta-\beta\gamma}f,
\end{equation*}
where $\alpha$, $\beta$, $\gamma$, $\delta$, $\mu_0$, $\mu_1$ and $\kappa$ are
constants; $\alpha$, $\beta$, $\gamma$, $\delta$ are defined up to a nonzero
multiplier, $\alpha\delta-\beta\gamma\neq0$ and $\kappa\neq0$. The generalized
equivalence group of the class~\eqref{GBE} coincides with the usual one.

A \emph{reduction operator} of an equation from the class~\eqref{GBE} is a vector field of the form
\begin{equation}\label{Q}
Q = \tau(t,x,u)\partial_t + \xi(t,x,u)\partial_x + \eta(t,x,u)\partial_u
\end{equation}
that leads to an ansatz reducing the initial equation to an ordinary differential equation. 
We carry out the nonclassical symmetry analysis of the class~\eqref{GBE}
following~\cite{popo08b,popo07a} and consider only reduction operators with $\tau\neq0$ ($\tau=0$ is
the ``no-go'' case for evolution equations~\cite{fush92,kunz08b,zhda98a}). 
Due to the equivalence relation among reduction operators we can
divide~$Q$ by $\tau$, so claim that $\tau=1$ in~\eqref{Q}.

The conditional invariance criterion (see~\cite{zhda99a}) implies the system of
determining equations with respect to the coefficients~$\xi$ and~$\eta$. Partially
solving this system, we derive expressions for~$\xi$ and~$\eta$, which are
polynomials of~$u$ with coefficients depending on $t$ and $x$,
\begin{equation*}
\xi = \xi^1(t,x)u + \xi^0(t,x), \\
\eta=\frac{\xi^1(\xi^1-1)}{3f}u^3 +
\left(\xi^1_x+\frac{\xi^1\xi^0}{f}\right)u^2+\eta^1(t,x)u+\eta^0(t,x).
\end{equation*}
Substituting these expressions into the system of determining equations and
splitting with respect to $u$ wherever it is convenient we obtain a system of
differential equations where $\xi^1$, $\xi^0$, $\eta^1$ and $\eta^0$ are assumed as
unknown functions,
\begin{gather*}
\xi^1(\xi^1-1)(2\xi^1+1)=0,
\\[.5ex]
-f_x(\xi^1)^2+f_x\xi^1+\xi^1\xi^0(2\xi^1+1)+4f\xi^1\xi^1_x=0,
\\
\frac{f_t}{f}(\xi^1-1)-2\frac{f_x}{f}\xi^1\xi^0-\frac{f_x}{f}\xi^0+3f\xi^1_{xx}
+2(\xi^1_x\xi^0+\xi^1\xi^0_x)+(2\xi^1+1)\eta^1-\xi^1_t+\xi^0_x=0,
\\
\xi^0_t+2\xi^0\xi^0_x+f\xi^0_{xx}-\eta^0(2\xi^1+1)=\frac{f_t}{f}\xi^0+\frac{f_x}{f}(\xi^0)^2+2f\eta^1_x,
\\
\eta_t+u\eta_x+f\eta_{xx}+2\xi_x\eta-\frac{f_t}{f}\eta-\frac{f_x}{f}\xi\eta=0.
\end{gather*}
We rewrite the last equation in terms of $\xi^1$, $\xi^0$, $\eta^1$ and $\eta^0$ and
split it with respect to $u$ only in particular cases.

The first determining equation obviously leads to the three possible cases depending on a value of~$\xi^1$: $\xi^1=1$, $\xi^1=-\frac12$ and $\xi^1=0$.

In the first case all
reduction operator coefficients except~$\xi^1$ vanish. The corresponding vector
field $Q=\partial_t+u\partial_x$ is a unique common reduction operator for all
equations from the class~\eqref{GBE}. The family of $Q$-invariant solutions consists
of the functions $u(t,x)=(x+c_1)/(t+c_2)$, where $c_1$ and $c_2$ are arbitrary
constants.

The second case is possible only if $f=1\bmod G^\sim$, 
i.e.\ we are faced with the problem of finding reduction operators with $\xi^1=-\frac12$ for the classical Burgers equation. 
It was established in~\cite{arri02a,mans99a} that 
solving the corresponding system of three differential equations in three unknown functions~$\xi^0$, $\eta^0$ and $\eta^1$
is equivalent to solving three copies of the linear heat equation. 
Therefore, it has been referred to as a ``no-go'' problem.

Consider the third case where $\xi=\xi^0(t,x)$. It is known from~\cite{arri93a} that
for the classical Burgers equation each reduction operator of this kind is
equivalent to a Lie symmetry operator. We prove that there are only two essential
subcases for~$f \neq \const$. In the first one, when $\xi^0_{xx}=0$, the reduction
operators have the form
\begin{equation*}
Q=\partial_t + \frac{(at+b_1)x+d_1t+d_0}{at^2+(b_1+b_2)t+c}\partial_x +
\frac{-(at+b_2)u+ax+d_1}{at^2+(b_1+b_2)t+c}\partial_u,
\end{equation*}
i.e.\ they are also equivalent to Lie symmetry operators. 
This paper is devoted to the other subcase, $\xi_{xx}\neq0$. 
It can be proved that then $\eta^1_t=(\eta^1)^2$ and $\eta^0=-\eta^1\xi^0$ 
and hence both $\eta^1$ and $\eta^0$ can be set to zero
using a transformation from the equivalence group~$G^\sim$.

We plan to present the proof and detailed consideration of all cases in a forthcoming paper.

\section{Connection with potential fast diffusion equations}\label{MAINsection}

Consider reduction operators of an equation of the form~\eqref{GBE} with $\xi_u=0$
while $\xi_{xx}\neq0$.
As mentioned above, under these conditions we have $\eta=0\bmod G^\sim$
and the system of determining equations is reduced to the system
\begin{gather}\label{deq1PFDEcase}
f_t+\xi f_x-\xi_xf=0,
\\\label{deq2PFDEcase}
\xi_t+\xi\xi_x+f\xi_{xx}=0,
\end{gather}
which is well determined as it consists of two differential equations in two unknown functions~$f(t,x)$ and~$\xi(t,x)$.
Moreover, this system is in Kovalevskaya form and hence it has no nontrivial differential consequences.

It is impossible to find the general solution of the system~\eqref{deq1PFDEcase}, \eqref{deq2PFDEcase}. 
In other words, we single out a ``no-go'' case in which one
cannot completely describe reduction operators of equations from the class~\eqref{GBE}. 
At the same time we are able to construct particular solutions of this system, 
which result in exact solutions of equations from the class~\eqref{GBE}
with special values of the arbitrary element~$f$. 
The above construction is realized via establishing a connection between the
system~\eqref{deq1PFDEcase}, \eqref{deq2PFDEcase} and a potential fast diffusion equation. 
For this purpose, we write equation~\eqref{deq1PFDEcase} in conserved
form, $(1/f)_t+(\xi/f)_x=0$, and then introduce the corresponding potential
$\theta=\theta(t,x)$, which is defined by the equations $\theta_t=\xi/f$ and
$\theta_x=-1/f$. Hence~$f$ and~$\xi$ can be expressed in terms of derivatives
of~$\theta$,
\begin{equation}\label{xi0andf}
f=-\frac{1}{\theta_x}, \quad \xi=-\frac{\theta_t}{\theta_x}.
\end{equation}
Substituting these expressions into the equation~\eqref{deq2PFDEcase} and factorizing, we obtain
\begin{equation*}
(\theta_x\partial_t-\theta_t\partial_x)
\left(\frac{\theta_t}{\theta_x}-\frac{\theta_{xx}}{(\theta_x)^2}\right)=0.
\end{equation*}
Integrated once and multiplied by~$\theta_x$, the last equation leads to
\begin{equation}\label{GFDE}
\theta_t=\frac{\theta_{xx}}{\theta_x}+h(\theta)\theta_x,
\end{equation}
where $h$ is an arbitrary smooth function of~$\theta$. Thus, for an
arbitrary solution $\theta$ of the equation~\eqref{GFDE} the vector field
$Q^\theta=\partial_t-(\theta_t/\theta_x)\partial_x$ is a reduction operator of the
equation of the form~\eqref{GBE} with $f=-1/\theta_x$.

It is impossible to solve the equation~\eqref{GFDE} for an arbitrary $h$ but
some interesting results can be obtained when
$h=\mu=\const$.
In this case the equation~\eqref{GFDE} is mapped by the transformation
$\tilde{t}=t$, $\tilde{x}=x+\mu t$ of the independent variables
to the potential fast diffusion equation
\begin{equation}\label{PFDE}
\theta_{\tilde{t}} = \frac{\theta_{\tilde{x}\tilde{x}}}{\theta_{\tilde{x}}},
\end{equation}
for which many exact solutions are known and all of them can be used in order to construct exact solutions of generalized Burgers equations.

As the Galilean boost $\tilde t=t$, $\tilde x=x+\mu t$, $\tilde u=u+\mu$, $\tilde f=f$,
where $\mu$ is an arbitrary constant, belongs to the equivalence group~$G^\sim$,
we can assume that $\mu=0\bmod G^\sim$.

The general equation~\eqref{GFDE} is mapped to the variable-coefficient potential fast diffusion equation
$\zeta_\tau=H(y)\zeta_{yy}/\zeta_y$ by a more complicated hodograph-type transformation
\begin{gather*}
\mbox{the new independent variables:}\qquad\tau=t, \quad y=g(\theta),
\\
\lefteqn{\mbox{the new dependent variable:}}\phantom{\mbox{the new independent variables:}\qquad }\zeta=x,
\end{gather*}
where $g(\theta)=\int e^{\int h\,d\theta}d\theta$ and $H(y)|_{y=g(\theta)}=e^{\int h\,d\theta}$. 
The corresponding fast diffusion equation is $v_\tau=(H(y)v_y/v)_y$. 
Variable-coefficient equations of the form $\zeta_\tau=H(y)\zeta_{yy}/\zeta_y$ 
are less studied than the constant-coefficient equation of the same form, i.e.\ equation~\eqref{PFDE}.
In particular, there are no exact solutions for these equations in the literature.

\section{Exact solutions of generalized Burgers equations}\label{LASTsection}

The ansatz $u=\varphi(\omega)$, $\omega=\theta(t,x)$ constructed with the operator~$Q^\theta$ reduces 
the generalized Burgers equation of the form~\eqref{GBE} with the value $f=-1/\theta_x$
to the ordinary differential equation $\varphi_{\omega\omega}-\varphi\varphi_\omega-h(\omega)\varphi_\omega=0$, 
which cannot be completely solved for a general value of~$h$.
At the same time, in the case~$h=\mu=\const$ the reduced equation is once integrated to the Riccati
equation $\varphi_\omega=\frac12\varphi^2+\mu\varphi+2\nu$, where $\nu$ is an integration constant.
It is remarkable that this is the same case when the corresponding equation~\eqref{GFDE} for~$\theta$ can be mapped to the potential fast diffusion equation~\eqref{PFDE}.
As the Riccati equation has constant coefficients it is obviously integrated using the substitution~$\varphi=-2\psi_\omega/\psi$.
Moreover, up to $G^\sim$-equivalence we can set $\mu=0$ from the very beginning.
As a result, we obtain the expression
\begin{equation}\label{SolFamily}
\varphi=\left\{
\begin{array}{ll}
\ds -2\varkappa\frac{c_1e^{\varkappa\omega}-c_2e^{-\varkappa\omega}}{c_1e^{\varkappa\omega}+c_2e^{-\varkappa\omega}}, & \nu<0,\\[2ex]
\ds -\frac{2c_2}{c_1+c_2\omega}, & \nu=0,\\[2ex]
\ds 2k\frac{c_1\sin k\omega-c_2\cos k\omega}{c_1\cos k\omega +c_2\sin k\omega}, & \nu>0,
\end{array}\right.
\end{equation}
where $\varkappa=\sqrt{-\nu}$, $k=\sqrt{\nu}$ and
$c_1$ and $c_2$ are arbitrary constants (only the ratio of these constants is essential).

This allows one to construct, for each known solution of the potential fast diffusion equation~\eqref{PFDE},
three families of solutions of the generalized Burgers equation with the value
$f=-1/\theta_x$ by substituting $\omega=\theta(t,x)$ into~\eqref{SolFamily}.
Note that in view of~\eqref{deq2PFDEcase} the function~$\xi=-\theta_t/\theta_x$
is also a solution of the same generalized Burgers equation with~$f=-1/\theta_x$.

The widest list of solutions of the potential fast diffusion equation~\eqref{PFDE} is presented
in~\cite{popo07a}.
They can be obtained, e.g., by integration of well-known solutions of the fast diffusion equation
(see ibid and also~\cite{gand01b,poly04book,pukh94book,qu99a,rose95a}).
These solutions together with the corresponding values of $f$ and $\xi$ are collected
in Table~\ref{TableSolutionOfGBE}.

The families of solutions constructed are extended to solutions of other generalized Burgers equations by transformations from the equivalence group~$G^\sim$.

\begin{table}[ht]
\footnotesize \caption{
Values of~$f$, $\xi$ and~$\theta$.
Here $\lambda$ is an arbitrary constant.}\label{TableSolutionOfGBE}
\begin{center}\renewcommand{\arraystretch}{1.3}
\begin{tabular}{|r|c|c|l|}
\hline
\hfil N &$f(t,x)                        $&$\xi(t,x)                    $&\hfil$\theta(t,x)                                                $\\
\hline%
 1&$\ds 1+e^{-t-x}                      $&$\ds e^{t+x}                 $&$\ds -\ln(e^t+e^{-x})                                            $\\
 2&$\ds -1                              $&$\ds 0                       $&$\ds x                                                           $\\
 3&$\ds 1-e^{-t-x}                      $&$\ds -e^{t+x}                $&$\ds -\ln|e^t-e^{-x}|                                            $\\
 4&$\ds -e^{-x}                         $&$\ds -e^{-x}                 $&$\ds e^x+t                                                       $\\
 5&$\ds t-x-\lambda t e^{-\frac{x}{t}}  $&$\ds 1-\lambda e^{-\frac xt} $&$\ds \ln|t|+\int{\frac{dw}{w-1+\lambda e^{-w}}}\,\bigg{|}_{w=x/t}$\\[1.1ex]
 6&$\ds -\frac{t^2-x^2}{2t}             $&$\ds \frac{x}{t}             $&$\ds \ln\left|\frac{x-t}{x+t}\right|                             $\\[1.1ex]
 7&$\ds -\frac{x^2}{2t}                 $&$\ds \frac{x}{t}             $&$\ds -\frac{2t}x                                                 $\\[1.1ex]
 8&$\ds -\frac{t^2+x^2}{2t}             $&$\ds \frac{x}{t}             $&$\ds 2\arctan\frac{x}{t}                                         $\\[1.1ex]
 9&$\ds -\frac{\cos^2x}{2t}             $&$\ds -\frac{\sin2x}{2t}      $&$\ds 2t\tan x                                                    $\\[1.1ex]
10&$\ds \frac{\cosh^2x}{2t}             $&$\ds -\frac{\sinh2x}{2t}     $&$\ds -2t\tanh x                                                  $\\[1.1ex]
11&$\ds -\frac{\sinh^2x}{2t}            $&$\ds \frac{\sinh2x}{2t}      $&$\ds -2t\coth x                                                  $\\[1.1ex]
12&$\ds \frac{\cos2x-\cos2t}{2\sin2t}   $&$\ds \frac{\sin2x}{\sin2t}   $&$\ds \ln\left|\frac{\sin(x-t)}{\sin(x+t)}\right|                 $\\[2.1ex]
13&$\ds \frac{\cosh2t-\cosh2x}{2\sinh2t}$&$\ds \frac{\sinh2x}{\sinh2t} $&$\ds \ln\left|\frac{\sinh(x-t)}{\sinh(x+t)}\right|               $\\[2.1ex]
14&$\ds \frac{\sinh2t-\sinh2x}{2\cosh2t}$&$\ds \frac{\cosh2x}{\cosh2t} $&$\ds \ln\left|\frac{\sinh(x-t)}{\cosh(x+t)}\right|               $\\[2.1ex]
15&$\ds \frac{\cosh2t+\cosh2x}{2\sinh2t}$&$\ds -\frac{\sinh2x}{\sinh2t}$&$\ds \ln\left|\frac{\cosh(x-t)}{\cosh(x+t)}\right|               $\\[2.1ex]
16&$\ds \frac{\cos2t-\cosh2x}{2\sin2t}  $&$\ds \frac{\sinh2x}{\sin2t}  $&$\ds 2\arctan(\cot t \tanh x)                                    $\\[1.4ex]
17&$\ds \frac{\cosh2t-\cos2x}{2\sinh2t} $&$\ds \frac{\sin2x}{\sinh2t}  $&$\ds 2\arctan(\coth t \tan x)                                    $\\[1.4ex]
\hline%
\end{tabular}
\end{center}
\end{table}

The values of~$\theta$ in Cases~12--17 of Table~\ref{TableSolutionOfGBE} are non-Lie solutions of the equation~\eqref{PFDE}. 
The maximal Lie invariance algebras of the corresponding generalized Burgers equations are zero, 
so it is clear that all exact solutions constructed for these equations are non-Lie ones.  
The other values of~$\theta$ presented in the table are Lie solutions of the equation~\eqref{PFDE} 
but only ansatz~\eqref{SolFamily} in Cases 2 and 6--8 and the values of~$\xi$ in all Cases 1--11 
provide Lie solutions of the corresponding equations of the form~\eqref{GBE}.

\section{Conclusion}

In the present paper the reduction operators of the form~\eqref{Q} with $\tau\neq0$ for the class~\eqref{GBE} of generalized Burgers equations are completely classified and exact solutions for certain equations of this kind are constructed.
(A more detail presentation of these and other results on symmetry analysis of generalized Burgers equations will appear soon.)
As far as we know, there are only a few exhaustive descriptions of nonclassical symmetries for important classes of nonlinear differential equations parameterized by arbitrary functions in the literature.
See, e.g., \cite{clar93b,arri94} for heat equations with nonlinear source, \cite{ivan10} for generalized Huxley equation and \cite{arri10} for systems of generalized Burgers equations.
The classification of reduction operators of equations from the class~\eqref{GBE} in the present paper gives one more example of such description.

The role of the equivalence group~$G^\sim$ is especially important in the course of this classification.
Applying equivalence transformations essentially simplifies both the computation and final results.

Another specific feature of this classification is the appearance of a no-go case which includes all nontrivial reductions operators.
There are no-go assertions on nonclassical symmetries of single differential equations \cite{fush92,kunz08b,popo08b,zhda98a}.
Similar results for classes of differential equations are less known.
If in a particular case the corresponding system of determining equations 
on unknown functional parameters involved in coefficients of reduction operators 
and arbitrary elements of the class under consideration is not overdetermined, 
we are not able to find all values of arbitrary elements for which equations 
from the class possess nontrivial sets of reduction operators.
At the same time, once a value of the tuple of arbitrary elements is fixed, 
the associated set of reduction operators may be completely defined.
The above situation arises for generalized Burgers equations from the class~\eqref{GBE}.
A vector field of the form~\eqref{Q} with $\tau=1$ and $\eta=0$ is a reduction operator 
of an equation of the class~\eqref{GBE} if and only if the coefficient~$\xi$ 
and the arbitrary element~$f$ satisfy the well-determined system~\eqref{deq1PFDEcase},
\eqref{deq2PFDEcase}.
This obstacle was partially overcome by means of the discovered connection between this system
and the potential fast diffusion equation~\eqref{PFDE}.
For each solution~$\theta$ of the potential fast diffusion equation~\eqref{PFDE}
we can construct at least three families of solutions of the generalized Burgers
equation of the form~\eqref{GBE} with $f=-1/\theta_x$.

\subsection*{Acknowledgements}
The authors thank Vyacheslav Boyko for useful discussions.
The research of ROP was supported by project P20632 of FWF.


\begin{thebibliography}{99}\itemsep=0ex
\footnotesize

\bibitem{ames72book}
Ames~W.F.,
{\it Nonlinear partial differential equations in engineering. Vol. II},
Academic Press, New York, 1972.

\bibitem{arri93a}
Arrigo~D.J., Broadbridge~P.\ and Hill~J.M.,
Nonclassical symmetry solutions and the methods of Bluman-Cole and Clarkson-Kruskal,
{\it J. Math. Phys.} {\bf 34} (1993), 4692--4703.

\bibitem{arri10}
Arrigo~D.J., Ekrut~D.A., Fliss~J.R.\ and Le~L.,
Nonclassical symmetries of a class of Burgers' systems.
{\it J. Math. Anal. Appl.} {\bf 371} (2010), 813--820.

\bibitem{arri02a}
Arrigo~D.J.\ and Hickling~F.,
On the determining equations for the nonclassical reductions of the heat and Burgers' equation,
{\it J. Math. Anal. Appl.} {\bf 270} (2002), 582--589.

\bibitem{arri94}
Arrigo~D.J., Hill~J.M.\ and Broadbridge~P.,
Nonclassical symmetry reductions of the linear diffusion equation with a nonlinear source,
{\it IMA J. Appl. Math.} {\bf 52} (1994), 1--24.

\bibitem{blum68th}
Bluman~G.W.,
{\it Construction of solutions to partial differential equations by the
use of transformation groups}, thesis, California Institute of Technology, Pasadena,
California, 1967.

\bibitem{blum69a}
Bluman~G.W.\ and Cole~J.D.,
The general similarity solution of the heat equation,
{\it J. Math. Mech.} {\bf 18} (1968/69), 1025--1042.

\bibitem{cate89a}
Cates~A.T.,
A point transformation between forms of the generalised Burgers equation,
{\it Phys. Lett.~A} {\bf 137} (1989), 113--114.

\bibitem{clar93b}
Clarkson~P.A.\ and Mansfield~E.L.,
Symmetry reductions and exact solutions of a class of nonlinear heat equations,
{\it Phys.~D} {\bf 70} (1994), 250--288. 

\bibitem{fush93book}
Fushchych~W.I., Shtelen~W.M.\ and Serov~N.I.,
{\it Symmetry analysis and exact solutions of equations of nonlinear mathematical physics},
Kluwer Acad. Publ., Dordrecht, 1993.

\bibitem{fush92}
Fushchych~W.I., Shtelen~W.M., Serov~M.I.\ and Popowych~R.O.,
$Q$-conditional symmetry of the linear heat equation,
{\it Dokl. Akad. Nauk Ukrainy} {\bf 12} (1992), 28--33.

\bibitem{gand01b}
Gandarias~M.L.,
New symmetries for a model of fast diffusion,
{\it Phys. Lett.~A} {\bf 286} (2001), 153--160.

\bibitem{ivan10}
Ivanova~N.M. and Sophocleous~C.,
On nonclassical symmetries of generalized Huxley equations,
in {\it Proceedings of 5th Int.\ Workshop "Group Analysis of Differential Equations and Integrable Systems" (Protaras, Cyprus, 2010)}, 2011, 91--98, arXiv:1010.2388.

\bibitem{katk65a}
Katkov~V.L.,
Group classification of solutions of the Hopf equation,
{\it Zh. Prikl. Mekh. i Tekhn. Fiz. (in Russian)} {\bf 6} (1965), 105--106.

\bibitem{king91c}
Kingston~J.G.\ and Sophocleous~C.,
On point transformations of a generalised Burgers equation,
{\it Phys. Lett.~A} {\bf 155} (1991), 15--19.

\bibitem{kunz08b}
Kunzinger~M.\ and Popovych~R.O.,
Singular reduction operators in two dimensions,
{\it J. Phys. A} {\bf 41} (2008), 505201, 24 pp., arXiv:0808.3577.

\bibitem{mans99a}
Mansfield~E.L.,
The nonclassical group analysis of the heat equation,
{\it J. Math. Anal. Appl.} {\bf 231} (1999), 526--542.

\bibitem{olve96book}
Olver~P.\ and Vorob'ev~E.M.,
Nonclassical and conditional symmetries, in R.L.~Anderson et al.,
{\it CRC handbook of Lie group analysis of differential equations. Vol. 3}, CRC, Boca Raton, FL, 1996, Chapter XI.

\bibitem{poly04book}
Polyanin~A.D.\ and Zaitsev~V.F.,
Handbook of nonlinear partial differential equations,
{\it Chapman \& Hall/CRC, Boca Raton, FL}, 2004.

\bibitem{popo08b}
Popovych~R.O.,
Reduction operators of linear second-order parabolic equations,
{\it J. Phys.~A} {\bf 41} (2008), 185--202, arXiv:0712.2764. 

\bibitem{popo10a}
Popovych~R.O., Kunzinger~M.\ and Eshraghi~H.,
Admissible transformations and normalized classes of nonlinear Schr\"{o}dinger equations,
{\it Acta Appl. Math.} {\bf 109} (2010), 315--359, arXiv:math-ph/0611061.

\bibitem{popo07a}
Popovych~R.O., Vaneeva~O.O.\ and Ivanova~N.M.,
Potential nonclassical symmetries and solutions of fast diffusion equation,
{\it Phys. Lett.~A} {\bf 362} (2007), 166--173, arXiv:math-ph/0506067.

\bibitem{pucc92a}
Pucci~E.,
Similarity reductions of partial differential equations,
{\it J. Phys.~A} {\bf 25} (1992), 2631--2640.

\bibitem{pukh94book}
Pukhnachov~V.V.,
Nonlocal symmetries in nonlinear heat equations,
{\it Energy methods in continuum mechanics (Oviedo, 1994)}, 75-99, {\it Kluwer Acad. Publ., Dordrecht}, 1996.

\bibitem{qu99a}
Qu~C.,
Exact solutions to nonlinear diffusion equations obtained by a generalized conditional symmetry method,
{\it IMA J. Appl. Math.} {\bf 62} (1999), 283--302.

\bibitem{rose95a} 
Rosenau~P.,
Fast and superfast diffusion processes, {\it Phys. Rev. Lett.} {\bf 74}
(1995) 1056--1059.

\bibitem{soph04a}
Sophocleous~C.,
Transformation properties of a variable-coefficient Burgers equation,
{\it Chaos Solitons Fractals} {\bf 20} (2004), 1047--1057.

\bibitem{wafo04d}
Wafo~Soh~C.,
Symmetry reductions and new exact invariant solutions of the generalized Burgers equation arising in nonlinear acoustics,
{\it Internat. J.~Engrg. Sci.} {\bf 42} (2004), 1169--1191.

\bibitem{whit74book}
Whitham~G.B.,
Linear and nonlinear waves. Pure and Applied Mathematics,
{\it Wiley-Interscience [John Wiley \& Sons], New York-London-Sydney}, 1974.

\bibitem{wood71th} 
Woodard~H.S.,
Similarity solutions for partial differential equations generated by finite and infinitesimal groups, Ph.D. Dissertation, University of Iowa, Iowa City,
Iowa, 1971.

\bibitem{wood71a} 
Woodard~H.S.\ and Ames~W.F.,
Similarity solutions for partial differential equations generated by finite and infinitesimal groups,
{\it Report. Subject Categories: theoretical mathematics; fluid mechanics}, 1971.

\bibitem{zhda98a}
Zhdanov~R.Z., Lahno~V.I.,
Conditional symmetry of a porous medium equation,
{\it Phys.~D} {\bf 122} (1998), 178--186.

\bibitem{zhda99a}
Zhdanov~R.Z., Tsyfra~I.M.\ and Popovych~R.O.,
A precise definition of reduction of partial differential equations,
{\it J. Math. Anal. Appl.} {\bf 238} (1999), 101--123, arXiv:math-ph/0207023.

\end{thebibliography}
\end{document}